# A new method for vehicle system safety design based on data mining with uncertainty modeling

Xianping Du[1], Binhui Jiang[2], Feng Zhu[2*]

[1] *Department of Mechanical and Aerospace Engineering, Rutgers University, Piscataway, NJ 08854, USA*

[2] *The State Key Laboratory of Advanced Design and Manufacturing for Vehicle Body, Hunan University, Changsha, 410082, China*

[3] *Hopkins Extreme Materials Institute and Department of Mechanical Engineering, Johns Hopkins University, Baltimore, MD, 21218, USA*

**Abstract:**

In this research, a new data mining-based design approach has been developed for designing complex mechanical systems such as a crashworthy passenger car with uncertainty modeling. The method allows exploring the big crash simulation dataset to design the vehicle at multi-levels in a top-down manner (main energy absorbing system – components - geometric features) and derive design rules based on the whole vehicle body safety requirements to make decisions towards the component and sub-component level design. Full vehicle and component simulation datasets are mined to build decision trees, where the interrelationship among parameters can be revealed and the design rules are derived to produce designs with good performance. This method has been extended by accounting for the uncertainty in the design variables. A new decision tree algorithm for uncertain data (DTUD) is developed to produce the desired designs and evaluate the design performance variations due to the uncertainty in design variables. The framework of this method is implemented by combining the design of experiments (DOE) and crash finite element analysis (FEA) and then demonstrated by designing a passenger car subject to front impact. The results show that the new methodology could achieve the design objectives efficiently and effectively. By applying the new method, the reliability of the final designs is also increased greatly. This approach has the potential to be applied as a general design methodology for a wide range of complex structures and mechanical systems.

## 1. Introduction

In recent years, the automotive industry has experienced the greatest demand to provide high safety rating vehicles with satisfactory performance of crashworthiness. In the event of a crash, the vehicle structure may undergo large plastic deformation to dissipate the kinetic energy. But likewise, it must sustain sufficient living space for its occupants. Vehicle safety requirements, such as Federal Motor Vehicle Safety Standard (FMVSS), New Car Assessment Program (NCAP), and Insurance Institute for Highway Safety (IIHS) have defined several specific performance requirements for vehicle crashworthiness [1-3]. To comply with these requirements, the vehicle structure must be designed to manage the crash energy effectively in the various crash modes likely to be encountered in fleet service. Many design factors, such as load path, structural deformations, and component collapse sequence, as well as vehicle type, size, weight, etc., should be considered simultaneously. To implement such a complicated system, efficient methods are necessary to enable the initiation of a sound design at the early concept design phase.

---

[*] Corresponding author: Hopkins Extreme Materials Institute, Johns Hopkins University, 3400 N Charles St, Baltimore, MD 21218, USA
E-mail: fengzhume@gmail.com (Dr. Feng Zhu); Xianping Du (xianping.du@rutgers.edu)

In the vehicle system, the energy-absorbing structures (EAS) are essentially designed as thin-walled beams or columns assembled to sustain plastic deformation during impact. The component thickness, geometry, and materials have a great effect on the performance of EAS [4-8]. It should be also noted that as a system, the interrelationship of different components must be considered as well. Since in the vehicle, EA components are integrated with bolts or welding, they have a complex coupling effect and cannot be deemed as independent of each other. It is quite often that if one component is modified, the others must be re-designed accordingly to ensure that the overall performance is still satisfactory [9, 10]. Such inherent coupling effect of components or design variables is complicated and usually implicit and "hidden" in the design data.

The conventional vehicle crashworthiness design methodologies are realized in a bottom-up manner, where each component is designed first and integrated to form the system-level design [9]. At the component level, the geometric parameters and materials are often used as the design variables [4, 11, 12]. The structural stiffness/intrusion or EA performance can be selected as the objective(s). The current CAE design approaches are based on a population of design alternatives and are widely employed in design optimization [7]. Within the design space, many design alternatives are created, and each one is termed as a design of experiment (i.e., DOE). These DOEs are computed with finite element analysis (FEA) subjected to the pre-defined constraints. Then optimization is conducted to find the best design through an evolutionary process. Typical optimization methods include Genetic Algorithm (GA), Simulated Annealing (SA), Evolutionary Algorithm (EA), Particle Swarm Optimization (PSO), etc. [13] After the component designs are completed, the components are integrated, and system level design starts. The EAS crash responses are evaluated via tests or simulations [14, 15]. If the performance is satisfactory, then the current design is accepted as the final design; Otherwise, go back to the component design and repeat this process until the satisfactory system level performance is achieved. Such approaches have been extensively applied in vehicle crashworthiness design. However, they have numerous limitations [9]:

1) The component design is implemented without prior knowledge from the system, such as loading and boundary conditions. Thus, the best component designs cannot guarantee good system performance. A lot of trial-and-error iterations are often needed.

2) As briefly mentioned above, conventional design approaches cannot reveal the interrelationship/coupling effect among different components implied in the simulation datasets. Since in complicated systems such as a vehicle, the components are integrated and have a coupling effect. Some components are "parents" whose behaviors could influence "children" components and then the overall response of the whole vehicle. In the design practice, the "parents" components must be determined first and then followed by the "children" components. Such intrinsic coupling effect of components or design variables is complicated and usually implicit and "hidden" in the vast amount of simulation data. Without this knowledge, it would be difficult to generate reasonable design and decision-making rules to tune each component in the right sequence.

3) It is difficult to link the detailed geometric variations of each component to the overall vehicle response. In the current system level vehicle safety design, the design variable is limited to the wall thickness of each beam or column, which alone is not able to describe the detailed profile of a component. In other words, the relationship between component level geometric variations and vehicle level response is not established. Since a huge number of variables are needed to accurately describe the geometry of a component, the inclusion of all these variables will significantly increase the number of DOEs and generally make the computational cost unaffordable.

To overcome these issues, a new data mining-based design approach has been developed for designing complex mechanical systems such as a crashworthy passenger car [16-20]. The method allows exploring the big crash simulation dataset to discover the underlying complicated relationships between response and design variables at multi-levels in a top-down manner (main energy-absorbing system-components-geometric features) and derive design rules based on the whole vehicle body safety requirements to make decisions towards the component and sub-component level design. Full vehicle and component simulation datasets are mined to build two decision trees. Based on the decision trees, the interrelationship among the design variables can be revealed and the design rules leading to a set of good designs can be derived. In a more recent work [9, 16], we extended this method by accounting for the uncertainty in the design dataset due to, for example, the manufacturing and computational errors [9]. A new decision tree algorithm for uncertain data (DTUD) was developed for evaluating the design performance variations due to the uncertainty in design variables, and the effectiveness of this new method was verified by designing a thin-walled energy-absorbing structure [16]. In the present paper, the DTUD approach is further extended to design a complicated mechanical system such as a crashworthy passenger car.

The remaining parts of this paper are organized as follows: In Section 2, the recently developed DTUD algorithm is reviewed. Based on this method, a framework for vehicle safety design is outlined. Section 3 formulates the current design objective and describes the baseline vehicle model to be used in the case study. Sections 4 and 5 present the design process and analyze the results, respectively. Additional discussions on the performance of this method are made in Section 6 and the findings of this work are summarized in Section 7.

## 2. Data mining method with uncertainty

*2.1 Data mining method for generating design rules*

The newly developed DMM (Data Mining Method) is a knowledge-based approach based on the binary classification decision tree technique [9, 19]. The decision tree is a tree-like diagram, which is composed of a root node on the top layer, several leaf nodes on the bottom layer, and the internal nodes between them. From the root node, each non-leaf node is a test for selecting an attribute [21] to maximize or minimize the value of a pre-defined attribute selection measure (ASM) [21] for *maximizing the label purity* of partitioned two subsets. The whole decision tree is generated by partitioning the training dataset recursively until reaching the termination criteria (for example, the maximum number of layers) with the outcomes labeled on the leaf nodes. Each path, connecting the root, internal, and leaf nodes, forms a branch. The design rules can be generated by interpreting each branch in a top-down sequence with an "if-then" decision-making logic.

An example to describe the concept is shown in Figure 1, where the failure risk of valve leaking is evaluated by the DMM with three parameters: tube thickness ($Tk$), internal diameter $D_e$ and inlet pressure ($P_e$). Learned from the simulation dataset, the generated decision tree contains two non-leaf nodes (elliptical blocks), three leaf-nodes (square blocks) with two labels (High Risk/HR and Low Risk/LR), and three branches (b1, b2, and b3). The parameters are determined by each non-leaf node for partitioning the current dataset into subsets with maximum purity and the partition criteria are shown on each path. By interpreting a selected branch, the design rule can be generated. For example, following b3, the design starts from the root node, where $P_e$ is tested. If $P_e \leq 1$ MPa, then go to the left and check $Tk$. If $Tk > 2$ mm, the label for the leaf node is LR, which means a low risk of leaking. Based on the tree, $P_e$ and $Tk$ are recognized as key variables and the range of their values can be determined. $De$ is not a key design variable since it is not included in the decision tree. The design rules based on this decision tree may help the designer better understand the design problem and make decisions efficiently.

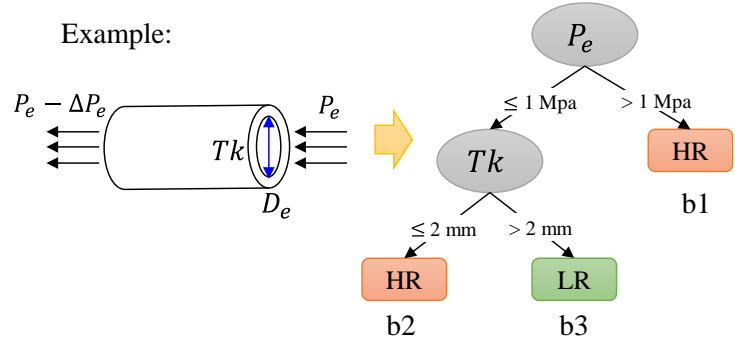

Figure 1 A simple example to demonstrate the application of a decision tree in the structural design

*2.2 A new decision tree for uncertain data (DTUD)*

In the vehicle crashworthiness problem, the uncertainty in the design data may be caused by the errors of numerical simulations, measurement, manufacturing, and material properties. The traditional decision tree algorithms cannot handle uncertain data. To resolve this issue, the DTUD was developed and verified by designing an energy-absorbing structure [9].

In this new algorithm, uncertain data are characterized by intervals with a pre-defined probability density function (PDF). Thus, no exact values are available as the split points like the traditional decision tree. $N$ (=10 in this example) points with uniform distribution in the design space are generated as the candidates of split points. For each non-leaf node, the attribute selection measure (ASM) is calculated on these points in the current dataset by selecting one point to split the current dataset binarily. In this study, the information gain ratio of a split point $S$ is used as the ASM and defined as:

$$\Omega(S) = \frac{\Pi(D) - \Pi_S(D)}{\Lambda(S)}, \tag{1}$$

where $\Pi(D)$ quantifies the label entropy of the current dataset $D$ as defined in Eq. (2); $\Pi_S(D)$ is the label entropy of the subsets of D after partitioned by $S$ as formulated in Eq. (3). $\Pi(D) - \Pi_S(D)$ is called the information gain. $\Lambda(S)$ is a normalization factor to avoid the bias of information gain to select the attribute with more outcomes as expressed in Eq. (4). For the current dataset, a $S$ value with the **largest** $\Omega(S)$ among all split points is used, which indicates the lowest information entropy after splitting.

$$\Pi(D) = \sum_{t=1}^{N_L} P^{Lt} \log_2(P^{Lt}), \tag{2}$$

$$\Pi_S(D) = \sum_{sp=1}^{N_v} \frac{|D_{sp}|}{|D|} \Pi(D_{sp}), \tag{3}$$

$$\Lambda(S) = \sum_{sp=1}^{N_v} \frac{|D_{sp}|}{|D|} \log_2\left(\frac{|D_{sp}|}{|D|}\right), \tag{4}$$

where $N_L$ is the number of label types (classes), and $P^{Lt}$ is the probability of the label $L_t$ in the current dataset (denoted as Label Probability (LP)) defined in Eq. (5). $N_v$ is the number of resulted partitions (=2, for the binary partition in this example). $\Pi(D_{sp})$ is the label entropy calculated for the *sp*th partition by Eq. (2). $|D|$ and $|D_{sp}|$ are the module or size of the current dataset and the $sp$th partition, respectively. For an uncertain dataset, $|D|$ can be calculated by Eq. (6).

$$P^{Lt} = \frac{\sum_{i=1}^{n_{Lt}} p_i^{t_{Lt}}}{|D|}, \tag{5}$$

$$|D| = \sum_{i=1}^{n_c} p_i^t, \tag{6}$$

where $n_c$ and $p_i^t$ are the total number of tuples (i.e., samples) in the current dataset $D$ and their probability in the current dataset, respectively. The number of $|D|$ be a relative robustness measure for a specific training dataset. Since a tuple or sample in the original training dataset is composed of intervals, this tuple is exactly a hypercube in the design space. The selected split point may split the hypercube into sub-cubes and assign them to different partitions. The weight or probability of the original tuple belonging to each partition, termed as Tuple Probability (TP), can be calculated by the integration of the joint PDF of this tuple on the corresponding sub-cube, that is, $p_i^t$, as introduced in [9]. $p_i^{t_{Lt}}$ denotes the TP of the $i$th tuple with label $L_t$ and $n_{Lt}$ is the number of tuples with label $L_t$ in the current dataset.

The DTUD construction algorithm is described in Figure 2. For the current dataset $D$, split points list ($SpliL$) can be generated by adding $N$ points on each interval and will be used as the candidates of splitting points.

---

**Algorithm**: Binary DTUD construction

**Input**: $D$: the current dataset

$SpliL$: Split points list

$Index$: Split point (Attribute) selection criterion

**Output**: a DTUD model

**Method**:

On the current node $CN$,

(1) if the tuples in $D$ have the same label or $SpliL$ is empty, then

  return $CN$ as a leaf node labeled with the LP of each class in $D$.

(2) else apply the $index$ for each element in $SpliL$ to find the best splitting point.

  for each of two outcomes, let $D_{sp}$ be the subset in $D$ satisfying outcome $sp$ ($sp$ = 1, 2);

  2.1) if $D_{sp}$ is empty, attach a leaf labeled with the LP of each class in $D$ to node $CN$;

  2.2) else attach the two child nodes returned by the splitting point to node $CN$;

(3) Update the $CN$ information: $D$, $SpliL$, by next non-leaf node.

  Return $CN$, its next-generation, and current information.

---

Figure 2 Basic algorithm for a binary DTUD construction.

As shown in Figure 3, a DTUD based on the pipeline example is grown in a top-down sequence in a similar recursive way with the traditional decision tree. However, the labels of the leaf nodes of DTUD are LP distributions instead of a specific category or value. In this example, branch b3 yields the HR and LR probabilities to be 0.04 and 0.96, respectively. If a label with the exact class rather than distribution is required for a leaf node, it can be assigned with the one having the highest LP value.

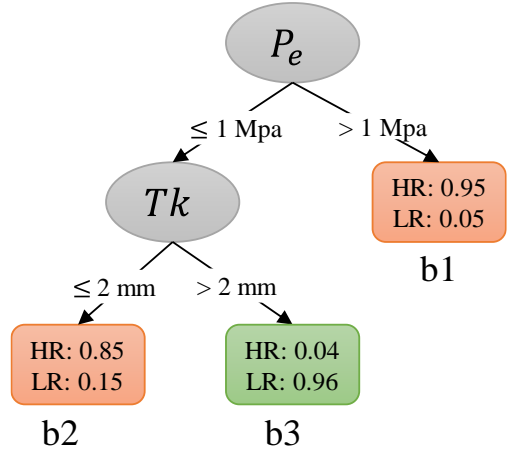

Figure 3 A DTUD for the pipeline problem

After the DTUD is built, the training accuracy of classification ($ACC_{tr}$) can be evaluated by:

$$ACC_{tr} = \frac{\sum_{i=1}^{n_l} p_i^c}{n_m}, \qquad (7)$$

where $n_l$ is the number of leaf nodes; $p_i^c$ is the TP summation of correctly classified tuples (with the same label as the leaf node) in $i$th leaf node. Using the trained DTUD, the new uncertain data with intervals can be classified with the LP distributions ($\vec{p_L}$) as

$$\vec{p_L} = \sum_{i=1}^{n_l} P_i^t \cdot \vec{p_i^L}, \qquad (8)$$

that is, the sum of the TP-weighted ($P_i^t$) LP distribution ($\vec{p_i^L}$) of each leaf node. The label with the maximum LP can be designated as the label of this tuple. In this way, the test accuracy can be calculated as the ratio of the correctly classified to the total number of samples in the test dataset.

Using the trained DTUD, the design rules can be generated following the same procedure in the traditional decision tree-based method. The LP distribution of each uncertain sample can be calculated by using Eq. (8). Under the uncertainty in design variables, identifying the samples with a low probability of preferred label can improve the reliability of the final design group. For the training and testing accuracy of DTUD and other details, please refer to [9].

*2.3 Framework for the systematic vehicle crashworthiness design methodology*

A framework for the vehicle full-frontal crashworthiness design with uncertainty in design variables is developed as shown in Figure 4. In this study, we focus on the shape design with no structural topology and material parameters considered. The framework can be decomposed into three main steps: (1) system level design; (2) component level design under uncertainty; and (3) design verification using a vehicle model.

*Step 1: System level learning*

The objective at the system level is to learn from simulation data and generate three decision-making rules to identify the following: $A_1$) *critical components* ($P_c$), $A_2$) *the feature response(s) ($\mathcal{F}_{P_c}$) of critical components* ($\mathcal{F}_{P_c}$), and $A_3$) *the boundary conditions of critical components* ($\mathcal{B}_{P_c}$). These identified rules are used as the prior knowledge for the component design to link the system and component levels. In this way, the design problem at the system level can be formulated as:

$$\begin{aligned}
\text{Determine:} \quad & \gamma_c = <P_c, \mathcal{F}_{P_c}, \mathcal{B}_{P_c}>, \\
\text{to maximize:} \quad & p('g'(\psi)|\gamma_c) \text{ and } \Phi('g'(\psi)|\gamma_c), \\
\text{subject to:} \quad & x_{vl} < x_v < x_{vu}, \\
& \mathbb{C}(x_v) \leq 0, \\
& \mathcal{L} = N_T ,
\end{aligned} \quad (9)$$

where $\gamma_c = <P_c, \mathcal{F}_{P_c}, \mathcal{B}_{P_c}>$ is the set of determined rules under the constraint of design variables $x_v$ between its lower ($x_{vl}$) and upper limit ($x_{vu}$) and constraint $\mathbb{C}(x_v)$. The number of decision tree layers $\mathcal{L}$ is also used as the model complexity constraint. $'g'$ represents the designs with good performance, which can be defined by the objective responses ($\psi$) of the design problem. The objective is to generate rules, $\gamma_c$ which maximizes the performance under the constraint with the probability ($p(\cdot)$) and robustness ($\Phi(\cdot)$). These are realized through the label purification process of the decision tree and representative branch selection, which will be discussed later.

In the step of the system level design, a validated whole vehicle model will be used to generate a simulation dataset. By running an initial crash simulation with this model, the main energy absorbing components can be identified, and they are the targets in the system-level design. In the space of virtual design variables (thicknesses of metal sheets in this study), a dataset of DOE is generated. $\mathcal{F}_p$ and $\psi$ of the design alternatives are quantified by the crash simulations. The training dataset is then generated by combining the $\mathcal{F}$ of each component and corresponding $\psi$. Learning from this dataset, the Rules $A_1$) and $A_2$) can be generated by the DMM along with the $A_3$) that is calculated by the crash simulation. All the information obtained will be transferred to Step 2.

*Step 2: Design at the component level*

In Step 2, the design objectives are to generate design rules for critical components to determine the following: $B_1$) *Critical design variables* and $B_2$) De*sign sub-space* for the response(s) of interest. Also, $\gamma_c$ from the system level is used as the prior knowledge. For each critical component, the design problem is formulated as:

$$
\begin{aligned}
&\text{Determine:} && x_p^s, \\
&\text{to maximize:} && p('g'(\psi_p)|x_p^s) \text{ and } \Phi('g'(\psi_p)|x_p^s), \\
&\text{subject to:} && x_{pl} < x_p < x_{pu}, \\
&&& \sigma_{x_p} = 0.1 \cdot x_p, \\
&&& \mathbb{C}_p(x_p, A_3) \leq 0, \\
&&& \mathcal{L}_p = N_T,
\end{aligned}
\tag{10}
$$

where $x_p$ denotes the design variables. This step is to find a set ($x_p^s$) of design alternatives or subspace of design variables. Thus, the probability and robustness, to generate designs with good performance, are maximized. Meanwhile, the constraints (design variables ($x_p$), model ($\mathbb{C}_p(\cdot)$), and decision tree layers ($\mathcal{L}_p$) and the 10% uncertainty level ($\sigma_{x_p}$) of design variables ($x_p$) are applied. The $\mathcal{F}_P$ requirement from system level ($A_2$) for each component can be incorporated into their responses $\psi_p$ or model constraint $\mathbb{C}_p$. $A_3$ will be used as the boundary and loading condition of the crash simulation at the component level. This could overcome the limitations of the traditional method by using roughly estimated or assumed initial velocity [22, 23] or constant velocity [24, 25] as the boundary conditions.

Based on this formulation and a component model parameterized with geometric design variables, a set of design alternatives is created, and their response is simulated through FEA to form the training set. Learned from this dataset, the rules ($B_{1-2}$) can be generated. To account for the uncertainty in design variables, DTUD is used in this step for the component design. In addition to the rule generation, the response variation due to uncertainty in design variables can be analyzed and used to improve the reliability of the final design set.

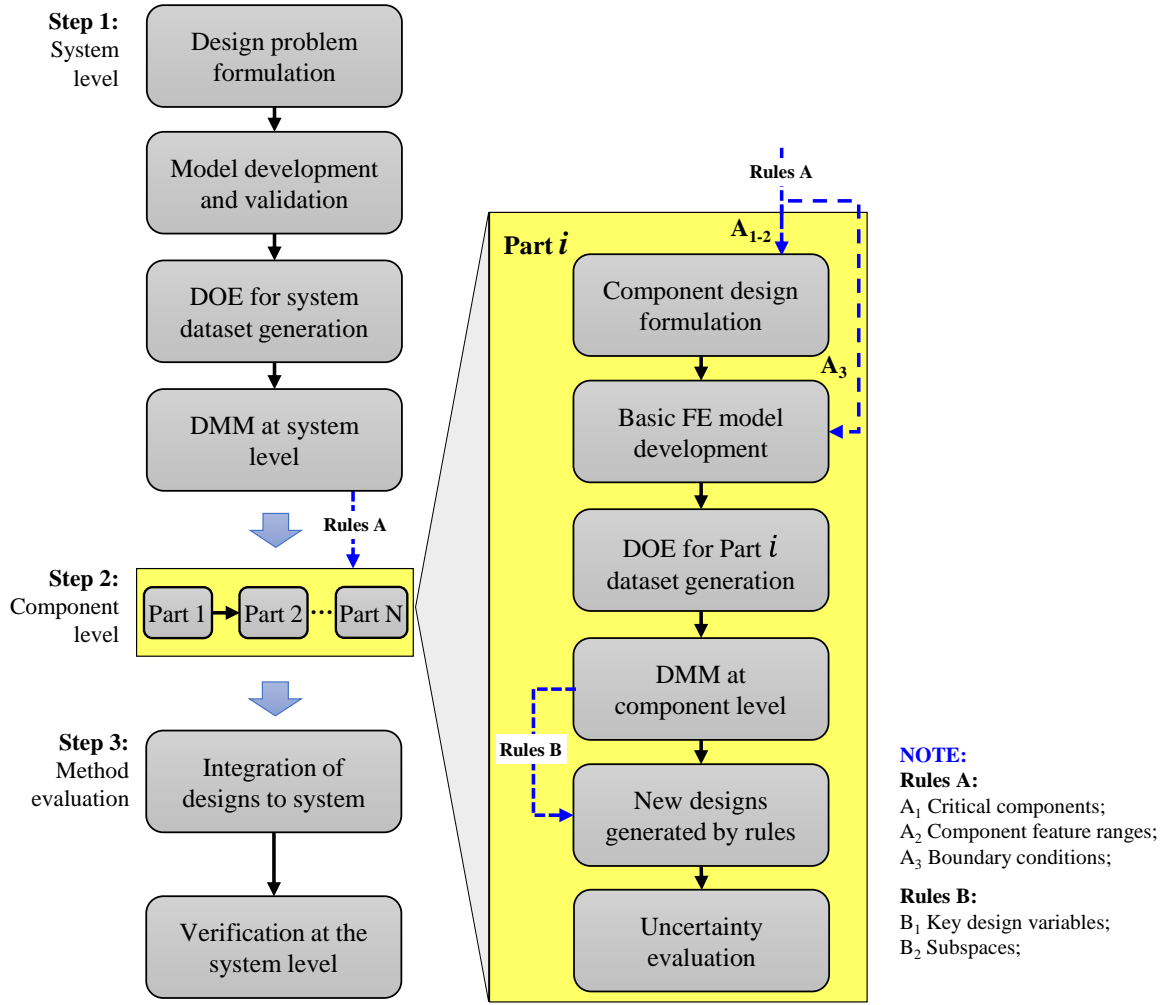

Figure 4 Implementation of the DMM on the vehicle crashworthiness design

*Step 3: Design verification using a vehicle model*

This step aims to verify the performance of the new method by conducting whole-vehicle simulations. The critical components designed through Steps 1 and 2 are integrated back to the whole vehicle model. Additional simulations are performed, and the model predictions, such as peak acceleration, energy absorption, and maximum intrusion, are compared with the results of the original design to verify the new method.

## 3. Baseline vehicle model and validation

*3.1 Basic model simplification*

In this section, a 2010 Toyota Yaris passenger car model is selected to implement the new design method described in Section 2. This model was developed by the former NCAC (National Crash Analysis

Center) of George Washington University through reverse engineering and is available in the NHTSA (National Highway Traffic Safety Administration) database (https://www.nhtsa.gov/es/crash-simulation-vehicle-models). The whole-vehicle model includes 974,383 elements, as illustrated in Figure 5(a). The response of this model has been validated by NCAC according to the NCAP full frontal impact conditions with an initial velocity of 56 km/h ($V_0$) as shown in Figure 5(a). The results are plotted in Figure 6. More information related to this vehicle model can be found on the NHTSA website [26].

Under the frontal impact condition, the rear side of the vehicle is less important since most of the crash energy is absorbed by the frontal structures. To reduce the computational cost, the model is simplified by removing elements at the rear side of the model. A mass block is added to the center of gravity of the removed parts to ensure that the inertia properties are not changed, as shown in Figure 5(b). After the simplification, the number of elements is reduced to 427,068. The simplified model is also validated by simulating two test cases (Nos. 05667 and 06221) available in the NHTSA vehicle crash test database (https://www-nrd.nhtsa.dot.gov/database/veh/veh.htm).

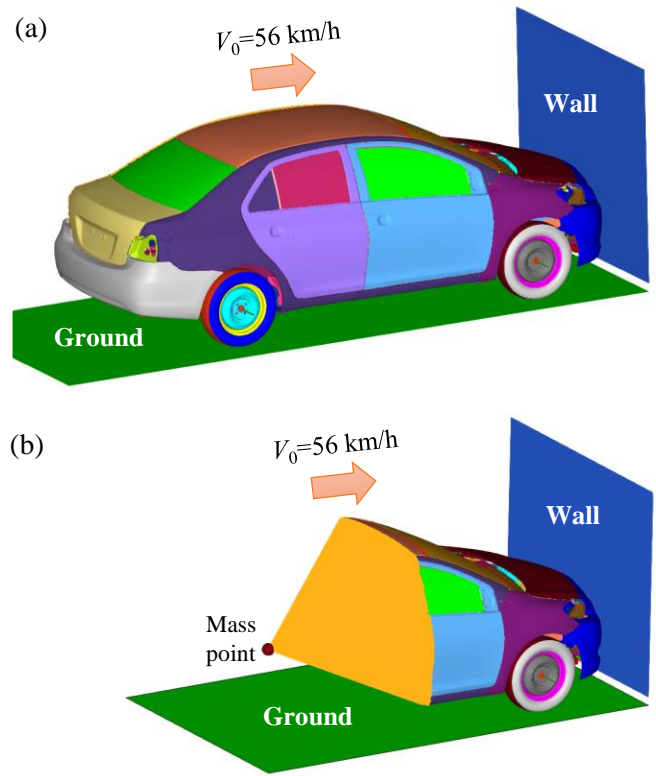

Figure 5 Finite element model of 2010 Toyota Yaris subject to a full-frontal impact specified in US NCAP: (a) the full vehicle model; (b) the simplified model

*3.2 Basic model validation*

Figure 6 shows the comparison of the simulated accelerations based on the original and simplified models and the test data available in the NHTSA database (i.e. No 05667 and 06221) at four locations. Since the rear seats were removed, the mass block acceleration was taken as the response for the rear seats. For Test No. 05667, the acceleration data at the engine bottom are not available and thus not included in the comparison [26]. The results show that the responses of the original and simplified are similar, and they show a reasonable match with the test data. Therefore, the simplified vehicle model is verified and will be used in the subsequent design.

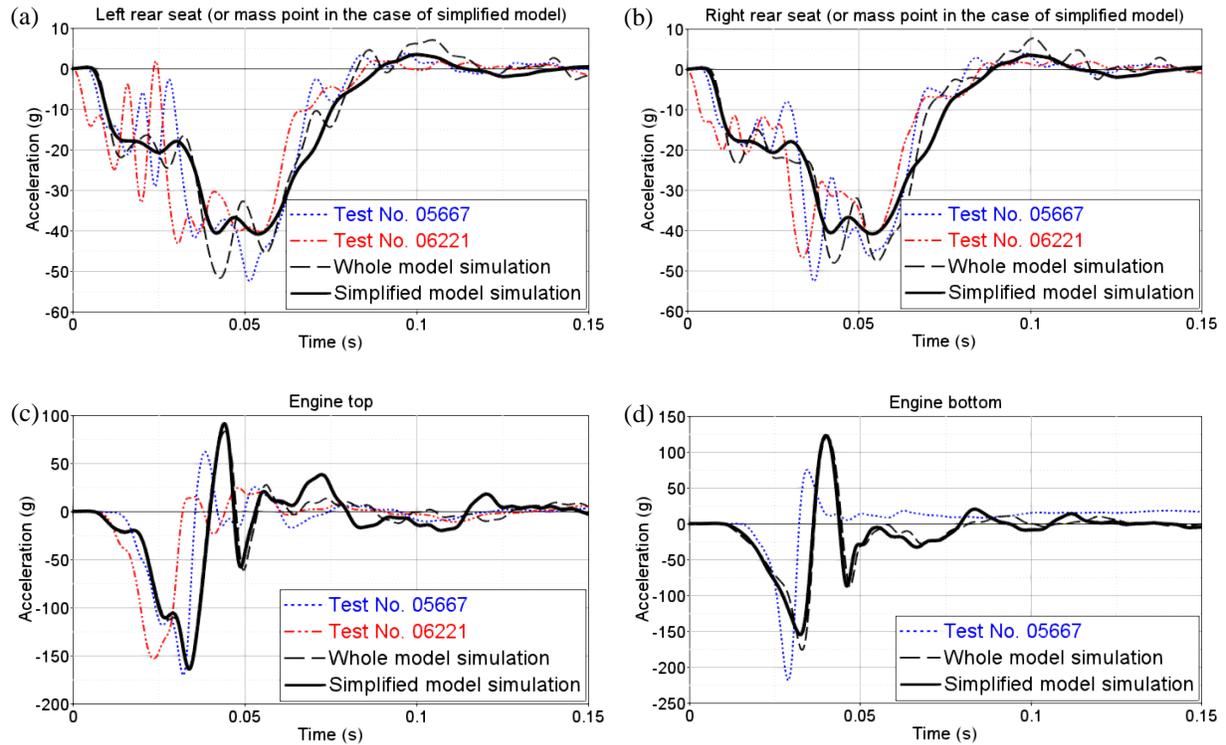

Figure 6 Validation of the original and simplified vehicle models under the condition specified in Figure 5.

## 4. Vehicle crashworthiness design using the new method

*4.1 System level design*

As briefly mentioned in Section 2.3, before the design, one initial simulation is used to screen the main energy-absorbing components. Based on the results, four components are identified and shown in Figure 7 since they absorb 85% of the impact energy. Each of them is composed of several metal sheets, that is, $P_1$ - bumper (①), $P_2$ - front frame (② + ③), $P_3$ - rail (④ + ⑤) and $P_4$ - floor support (⑥ + ⑦ + ⑧). To characterize the response of each component, here we define a parameter called "average stiffness ($avgstiff$)", which can be calculated in Eq. (11). This parameter is used as the attribute of components, $\mathcal{F}_P$, since it is closely related to energy-absorption, acceleration, and intrusion. Here it is used as the input feature of the decision tree.

$$avgstiff(P_i) = \frac{\bar{F}}{d_{P_i}} = \frac{\int_0^{d_{P_i}} F(u)\,du}{d_{P_i} \cdot d_{P_i}}, \tag{11}$$

where $\bar{F}$ and $d_{P_i}$ are the mean crush force and final deflection of $P_i$, respectively. $F(u)$ is the force-displacement history. To adjust the value of $avgstiff$, one can change the wall thickness of each component, its dimensions, geometric or material. In this study, to simplify the problem, only wall thickness is changed.

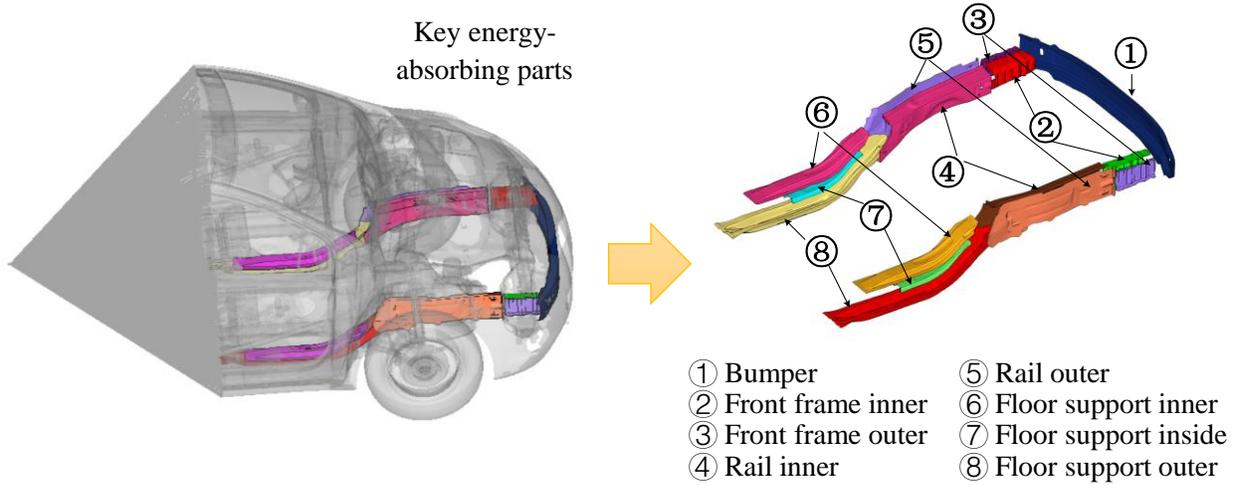

① Bumper    ⑤ Rail outer
② Front frame inner    ⑥ Floor support inner
③ Front frame outer    ⑦ Floor support inside
④ Rail inner    ⑧ Floor support outer

Figure 7 Four main energy absorbing components identified in the initial simulation (left) and their detailed structures (right)

Three design objectives ($\psi_1$, $\psi_2$ and $\psi_3$) are the peak crush force ($F_p$), firewall intrusion ($S_p$), and the total mass of the main EA components ($\psi_3$), which are expressed as

$$\psi_1 = F_p = \max(F(t)), \tag{12}$$

$$\psi_2 = S_p = \max(\text{avg}(s_1(t_k), s_2(t_k), s_3(t_k), s_4(t_k))), \tag{13}$$

$$\psi_3 = M = \sum_{Pi=1}^{Np} M_{Pi}, \tag{14}$$

where $F(t)$ is the time history of the crushing force and $\text{avg}(s_1(t_k), s_2(t_k), s_3(t_k), s_4(t_k))$ is the averaged intrusion of four pre-selected node markers on the fire-wall at the $k$th time step ($t_k$) with their locations shown in Figure 8. These markers are selected around the maximum deformation areas of the fire-wall in simulation. $S_p$ is the maximum among all time-steps. $M_{Pi}$ is the mass of the $P_i$. Thus, $\psi(x_v, V_0) = \langle F_p, S_p, M \rangle$ is defined. The system level design is to determine the components' feature response, $avgstiff$ to ensure a high crashworthiness performance ($\psi$). The design problem at the system level (Eq. (9)) can be re-written in the form of Eq. (15).

$$\begin{aligned}
\text{Find the rule:} \quad & \gamma_c = <P_c, avgstiff(P_c), \mathcal{B}_{P_c}>, \\
\text{to maximize:} \quad & p('g'(\psi)|\gamma_c) \text{ and } \Phi('g'(\psi)|\gamma_c), \\
\text{subject to:} \quad & 2.0 \le T_{1,6,7} \le 3.0, \\
& 1.5 \le T_{2,4,5,8} \le 2.5 \text{ mm}, \\
& 1.0 \le T_3 \le 2.0 \text{ mm}, \\
& V_0 = 56 \text{ km/h}, \\
& \mathcal{L}(x_v) = 6.
\end{aligned} \quad (15)$$

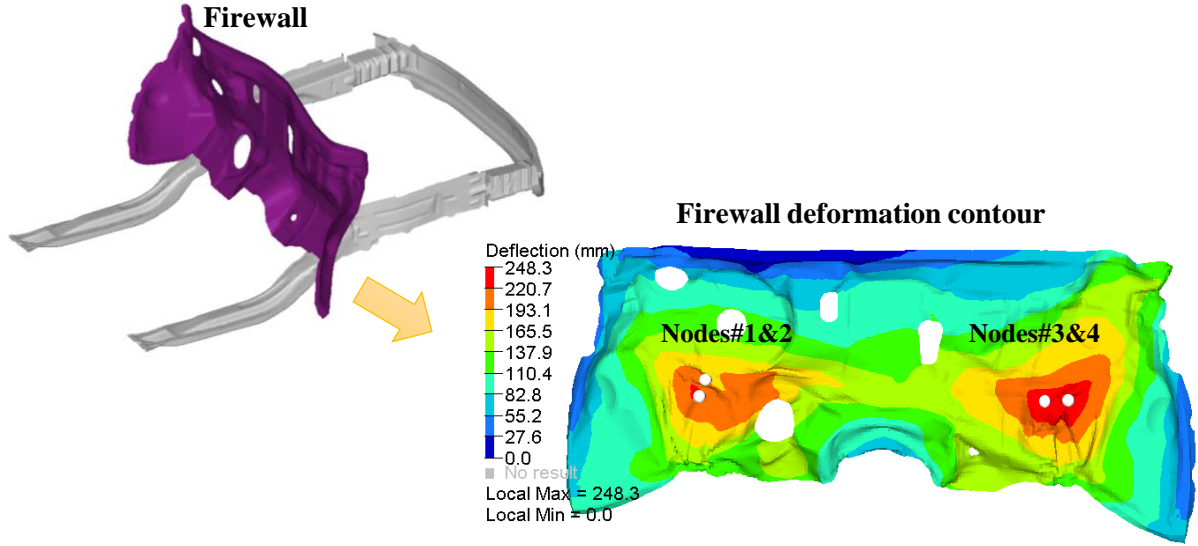

Figure 8 The location of the firewall and its deformation contour at the final timestep of the crash simulation

To generate the training dataset for system level design, 150 combinations of the different values of design variables are generated by Latin Hypercube Sampling (LHS). Previous studies indicated that for a design problem with $N$ independent parameters, $3N$ design instances should be sufficient to describe the design space [27, 28]. In the current system level design problem, there are eight thickness values applied as the design variables. Therefore, the aforementioned 3N rule is satisfied. The 150 FE models are computed, and the simulation results form a design dataset.

In the dataset, the distributions of the ranges of $avgstiff$ (unit: kN/m) are shown in Figure 9: $1{,}848.1 \le avgstiff(P_1) \le 10{,}243.7$, $382.0 \le vgstiff(P_2) \le 949.2$, $178.6 \le avgstiff(P_3) \le 988.2$, and $120.5 \le avgstiff(P_4) \le 296.1$. The labeling strategy is set as follows: good or "g": $F_p < 800$ kN, $S_p < 220$ mm and $M < 27$ kg; poor or "p": $F_p \ge 800$ kN, $S_p > 260$ mm and $M > 28$ kg; and intermediate or "m": the others. Thus, 21%, 81%, and 48% of samples are labeled as 'g', 'm', and 'p', respectively. It should be noted that the labeling or classification strategy in a particular design problem is often determined in an arbitrary way, and it can be adjusted based on the designer's experience or preference [references].

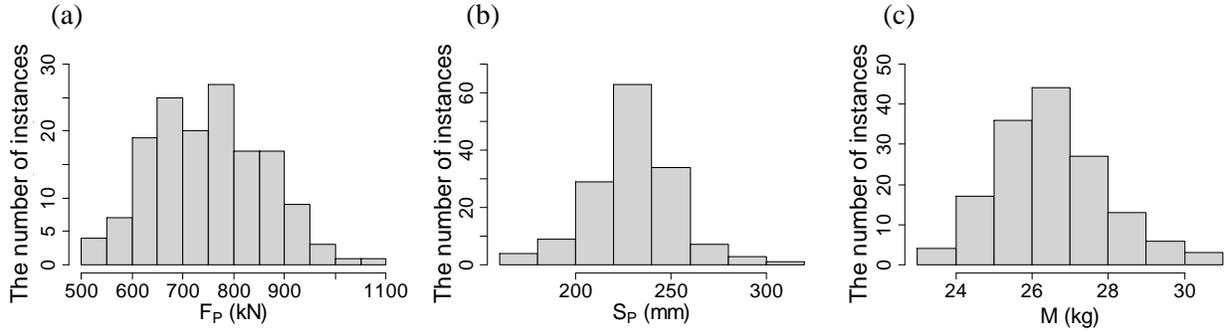

Figure 9 Distribution of the 150 samples in terms of three design objectives: (a) $F_P$, (b) $S_P$, and (c) $M$

A binary decision tree constructed based on the training dataset is shown in Figure 10. At the system level, no uncertainty is considered since this is not a detailed design stage. The mini-batch size of 30 and 5-fold cross-validation is used to improve the model accuracy to learn from this relatively small dataset. The result indicates that 88% of samples are correctly classified and assigned as their real labels. This tree is composed of 13 non-leaf (for average stiffness), 14 leaf nodes (for labels), and 14 branches (b1-b14). By following a branch, a design rule can be generated to reach the corresponding outcome (response label) on its leaf node. In Figure 10, branch b4 is selected as the representative of "g" due to its both high 'g' probability (7/8) and relatively more tuples (8) included for the high robustness. Three critical components (Rule $A_1$: $P_2$, $P_3$, and $P_4$) are identified. The subspace of the $avgstiff$ (Rule $A_2$) is generated: $487.3 \leq avgstiff(P_2) \leq 594.3$, $178.6 \leq avgstiff(P_3) \leq 404$, and $175.3 \leq avgstiff(P_4) \leq 296.1$. Also, the loading and boundary conditions of each component (Rule $A_3$) are derived from the simulations with the simplified vehicle model, which are detailed in the next section.

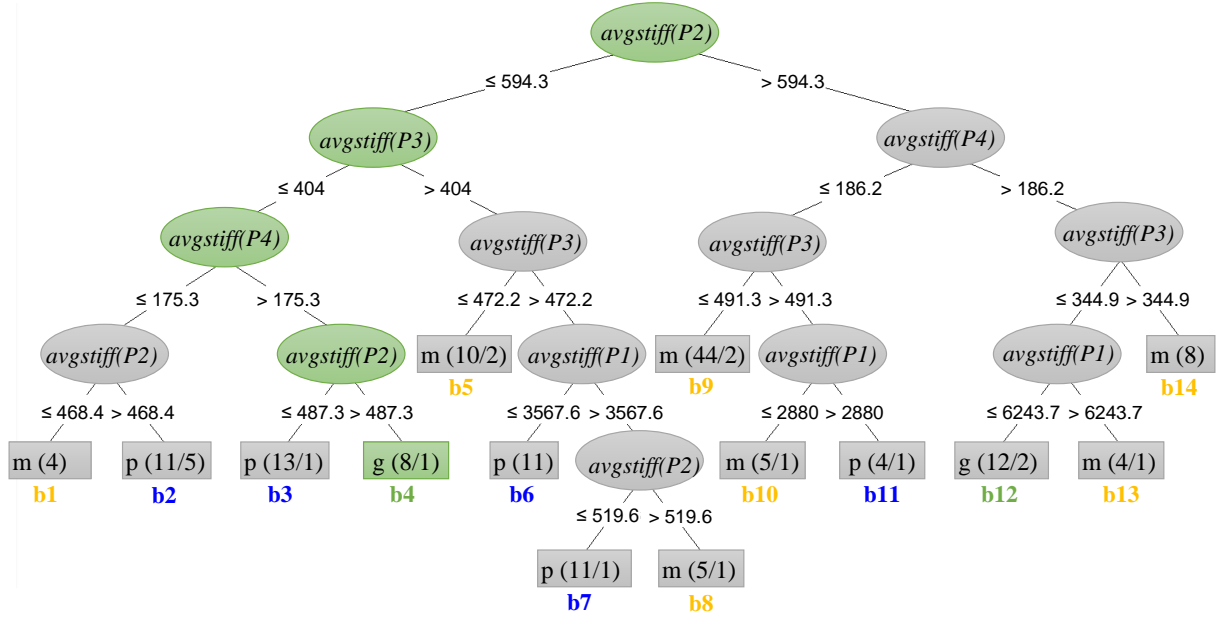

Figure 10 System-level decision tree. The denotation of the leaf node: Label (the number of designs in this leaf/the number of designs incorrectly classified)

*4.2 Component level design*

Based on the knowledge obtained from the system level, i.e. design rules ($A_1$, $A_2$, and $A_3$) and boundary conditions, detailed design for the selected components is conducted in this section.

4.2.1 Component parameterization and design objectives definition

In this step, the components are represented with a number of geometric parameters or design variables and then converted to the FE model. By adjusting the values of these design variables, the profile of the part can be modified. The node-based morphing method can be used to change their mesh based on the values of design parameters [29, 30]. In this method, the morphing equation is established between the predefined control points and nodes of finite element models. By changing the locations of predefined control points, the FE nodes are moved to new locations accordingly based on the morphing rules. This method has been applied widely in various engineering designs [29-32].

The components in the simplified vehicle model described in Section 3 (i.e., $P_2$, $P_3$, and $P_4$ shown in Figure 11(a)) are used as the baseline for geometric modification. Their typical deformation and energy dissipation modes are shown in Figure 11b~d, respectively.

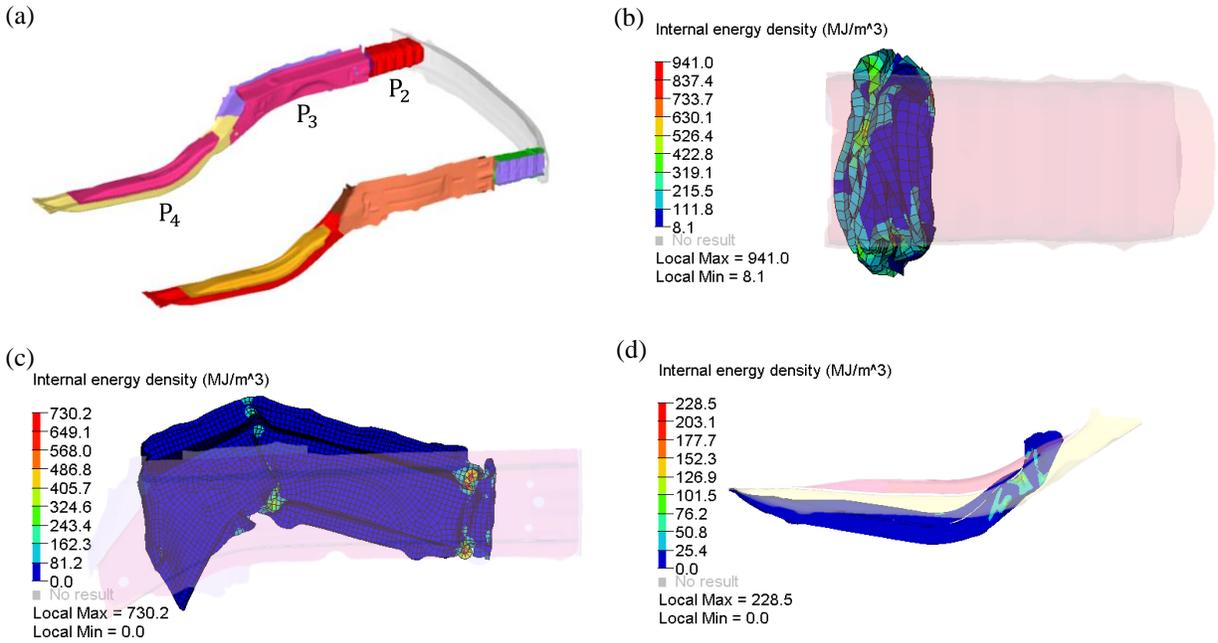

Figure 11 Three key energy-absorbing structures and their deformation modes as well as internal energy density distribution: (a) assembly of the three structures; (b) $P_2$; (c) $P_3$; and (d) $P_4$

The geometric parameters for the three selected critical components are determined and shown in Figure 12. On each cross-section, several morphing control points (MCPs) are placed at the corners due to their large curvature. The black MCPs are fixed and green MCPs are movable on their cross-section plane along the denoted axes ($y$ and/or $z$) in the cartesian coordinate system. Changing the locations of the MCPs, a mapping rule from the original to the new coordinates of MCPs is derived. Using this rule, the FE nodes of the baseline model can be morphed to the new MCP locations to generate a new mesh. The detailed morphing algorithm is described in Appendix A.

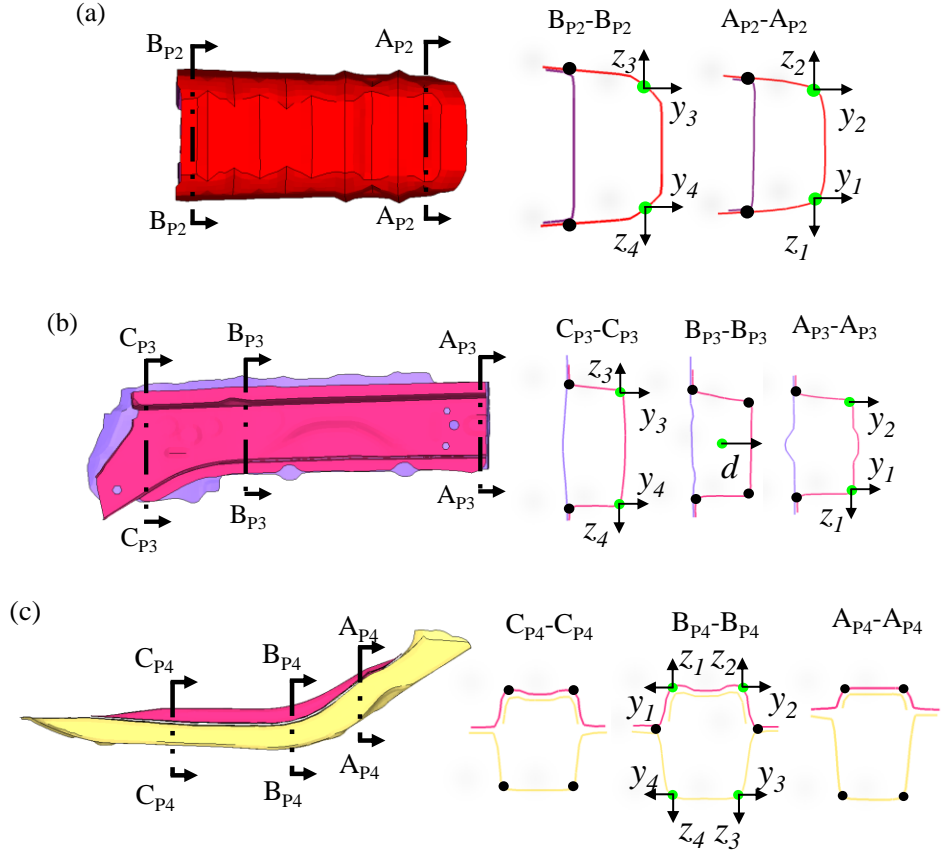

Figure 12 Geometric control points on the critical components (a) $P_2$, (b) $P_3$, and (c) $P_4$, where the black control points are fixed but the green points can be moved along the arrows on the axes.

The moving distances of the MCPs from the initial positions in the baseline model together with the thicknesses of the metal sheets are defined as the design variables and shown in Figure 7. At the component level design, specific energy ($SEA$) as defined in Eq. (16) is used as the design objective. For the lightweight purpose, the component mass ($M_P$ in Eq. (14)) is taken as the second objective.

$$SEA_P = \frac{\int_0^{d_P} F(u)du}{M_P}. \tag{16}$$

The component level design problem formulated in Eq. (10) is quantified for P2, P3, and P4 in Eqs. (17)-(19), respectively. To determine the correct boundary conditions for these components, the time-displacement history from $A_3$ at the system level design are applied.

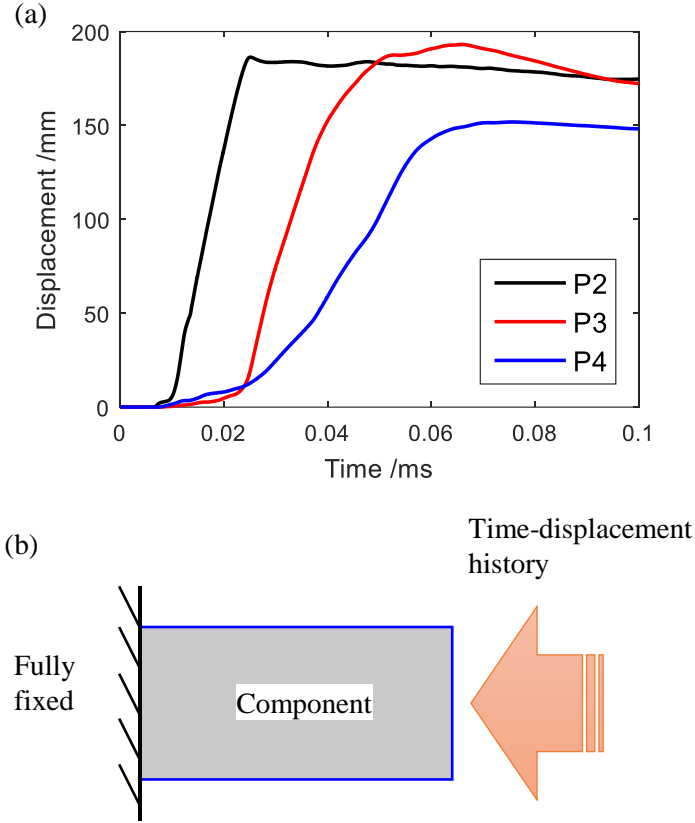

Figure 13 (a) The time displacement history of critical components from the system-level simulation and (b) the boundary conditions for the component level simulation

For **P$_2$**:
Find: $x_{P2}^S$,
to maximize: $p('g'(\psi_{P_2}))|x_{P2}^S)$ and $\Phi('g'(\psi_{P_2}))|x_{P2}^S)$,
subject to:
$1.5 \leq T_2 \leq 2.5$ mm,
$1.0 \leq T_3 \leq 2.0$ mm,
$-15 \leq P_{2\_y1,2,3,4} \leq 15$ mm, (17)
$-20 \leq P_{2\_z1,2,3,4} \leq 20$ mm,
$487.3 < avgstiff(P_2) \leq 594.3$ kN/m,
$\sigma_{x_{p2}} = 0.1 \cdot x_{p2}$,
$\mathcal{L}_{P2} = 9$,

For **P$_3$**:
Find: $x_{P3}^S$,
to maximize: $p('g'(\psi_{P_3}))|x_{P3}^S)$ and $\Phi('g'(\psi_{P_3}))|x_{P3}^S)$, (18)

subject to:  $1.5 \leq T_{4,5} \leq 2.5$ mm,

$-5 \leq d \leq 5$ mm,

$-15 \leq P_{3\_y1,2,3,4} \leq 15$ mm,

$-20 \leq P_{3\_z1,3,4} \leq 20$ mm,

$178.6 \leq avgstiff(P_3) \leq 404$ kN/m,

$\sigma_{x_{p3}} = 0.1 \cdot x_{p3}$,

$\mathcal{L}_{P3} = 9$,

For $\mathbf{P_4}$:

Find: $x_{P4}^S$,

to maximize: $p('g'(\psi_{P_4}))|x_{P4}^S)$ and $\Phi('g'(\psi_{P_4}))|x_{P4}^S)$,

subject to: $2.0 \leq T_{6,7} \leq 3.0$ mm,

$1.5 \leq T_8 \leq 2.5$ mm,

$-15 \leq P_{4\_y1,2,3,4} \leq 15$ mm, (19)

$-15 \leq P_{4\_z1,2,3,4} \leq 15$ mm,

$175.3 < avgstiff(P_4) \leq 296.1$ kN/m,

$\sigma_{x_{p4}} = 0.1 \cdot x_{p4}$,

$\mathcal{L}_{P4} = 9$,

where $x_{Pi}^S$ ($i = 2, 3,$ and $4$) is the critical variable and subspace of the $i$th component. $'g'$ is the label for good designs, which is specified by the requirement of $SEA_P$ and $M_P$ as listed in Table 1.

4.2.2 Generation of uncertain datasets

For each component, 300 design alternatives are created by LHS considering the number of design variables, degree of nonlinearity, and time cost [27, 28, 33]. All these DOEs are converted to FE models and simulated. A dataset is generated for each component including the combinations of design variable values and corresponding $SEA_P$ and $M_P$. The datasets are labeled using the corresponding labeling strategy specified in Table 1. After the labeling, the distribution of these three classes in each dataset is also listed in Table 1. The newly generated simulation datasets are extended by adding uncertainty into the design variables. A range is defined for each design variable, and the original exact values are used as the means with a deviation R=10%. Considering the random nature of the errors in simulations, measurement, or manufacturing, Gaussian distribution is assumed on each interval ($[A_{ij}^{CU}\ A_{ij}^{CL}]$) as expressed in Eq. (20)

$$f(A_{ij}) = C \cdot N(A_{ij}^m, (\frac{(A_{ij}^{CU} - A_{ij}^{CL})/2}{3})^2), \quad (20)$$

where $N(\cdot)$ is a gaussian distribution by taking the original exact value as the mean with the 3-sigma probability applied on the interval. C = 1/0.997 is a normalizing coefficient to adjust the probability on the interval to 1. For each critical component, their design variables are independent of each other, so the joint probability distribution of design variables is a truncated Gaussian distribution. In this way, the label of each sample is not changed after involving the uncertainty.

*Table 1 Labeling criteria for the three critical components and the class distribution in their corresponding training dataset*

| Component | $P_2$ | $P_3$ | $P_4$ |
|---|---|---|---|
| Labeling criteria | **'g'**: $SEA \geq 20{,}500$ J/kg, and $M_2 \leq 0.95$ kg<br>**'p'**: $SEA < 19{,}500$ J/kg, or $M_2 > 1$ kg<br>**'m'**: others | **'g'**: $SEA \geq 2{,}800$ J/kg, and $M_2 \leq 4.2$ kg<br>**'p'**: $SEA < 2500$ J/kg, or $M_2 > 4.7$ kg<br>**'m'**: others | **'g'**: $SEA \geq 525$ J/kg, and $M_2 \leq 5.5$ kg<br>**'p'**: $SEA < 475$ J/kg, or $M_2 > 6$ kg<br>**'m'**: others |
| Class distribution | 'g': 18%; 'm': 45%; 'p': 37%; | 'g': 16%; 'm': 60%; 'p': 23%; | 'g': 22%; 'm': 48%; 'p':29%; |

4.2.3 Generation of design rules at the component level

Using the datasets with uncertainty, DTUDs are generated in Figure 14 (Tree 1), Figure 15 (Tree 2), and Figure 16 (Tree 3) for $P_2$, $P_3$ and $P_4$, respectively. To achieve the design objective, a branch with compromising 'g' probability and robustness is selected as the design rules ($B_1$ and $B_2$). The branch b3 in Tree 1, b19 in Tree 2, and b19 in Tree 3 are selected from the corresponding tree for $P_2$, $P_3$ and $P_4$, respectively. These design rules for generating 'g' designs are summarized in Table 2. Using these rules, a set of design alternatives with 'g' label can be generated efficiently. For each of the critical components, 20 new design alternatives (not included in the training dataset) are generated by LHS within the reduced design space determined by the DYUD. These new designs will be used to evaluate the system level design, and their details are listed in Appendix B.

*Table 2 Design rules generated from DTUD for the critical components with a "g" label*

| Critical component | $P_2$ (**b3**) Tree1 | $P_3$ (**b19**) Tree2 | $P_4$ (**b19**) Tree3 |
|---|---|---|---|
| Rules | $-15 \leq P_{2\_y_1} \leq 5.80$ mm<br>$-15 \leq P_{2\_y_2} \leq -6.51$ mm<br>$-11.50 < P_{2\_y_3} \leq 15$ mm<br>$-15 \leq P_{2\_y_4} \leq 2.95$ mm | $1.5 \leq T_4 \leq 1.81$ mm<br>$-0.095 < d \leq 5$ mm<br>$-15 \leq P_{3\_y_1} \leq 6.10$ mm<br>$-12.21 \leq P_{3\_y_2} \leq 12.21$ mm<br>$-15 \leq P_{3\_y_3} \leq 8.65$ mm | $1.5 \leq T_8 \leq 2.12$ mm<br>$0.51 < P_{4\_z_1} \leq 15$ mm<br>$-15 < P_{4\_z_2} \leq -10.70$ mm<br>$-1.03 < P_{4\_z_3} \leq 15$ mm<br>$-13.89 < P_{4\_z_4} \leq 15$ mm |

Figure 14 Uncertain decision tree for Part 2 at the component level

Figure 15 Uncertain decision tree for Part 3 at the component level

Figure 16 Uncertain decision tree for Part 4 at the component level

4.2.4 Screening the samples with performance deterioration due to uncertainty by DTUD

To qualify the effect of the uncertainty on the component response, each uncertain sample in the uncertain design dataset is tested by corresponding DTUT in Figure 14-Figure 16 and the calculated LP values are listed in Table 3. The nominal LP of the corresponding DTUD branch that generates the design dataset is also included for comparison. As verified in our previous study [16], if the DTUD predicted 'g' probability is lower than the branch nominal LP greatly, the performance of this design alternative exhibits low reliability under uncertainty. For each dataset, all of the design alternatives are sorted based on their LP values. The 10 cases with the highest LP values (renumbered as No. 1-10) are selected and accepted as the final designs of each component since they have the highest stability.

*Table 3 LP distribution predicted by the DTUD under the uncertainty for the three critical components corresponding to the design alternatives in Appendix B (Sorted in descending order of predicted 'g' LPs)*

| No. | P2 (p('g')=0.89) | | | P3 (p('g')=0.87) | | | P4 (p('g')=0.92) | | |
|---|---|---|---|---|---|---|---|---|---|
| | p | m | g | p | m | g | p | m | g |
| **1** | 0.05 | 0.00 | **0.95** | 0.00 | 0.13 | **0.87** | 0.00 | 0.06 | **0.94** |
| **2** | 0.05 | 0.00 | **0.95** | 0.00 | 0.13 | **0.87** | 0.00 | 0.08 | **0.92** |
| **3** | 0.04 | 0.01 | **0.95** | 0.00 | 0.13 | **0.87** | 0.00 | 0.08 | **0.92** |
| **4** | 0.00 | 0.06 | **0.94** | 0.00 | 0.13 | **0.87** | 0.00 | 0.08 | **0.92** |
| **5** | 0.01 | 0.10 | **0.89** | 0.00 | 0.13 | **0.87** | 0.00 | 0.08 | **0.92** |
| **6** | 0.01 | 0.10 | **0.89** | 0.00 | 0.13 | **0.87** | 0.00 | 0.08 | **0.92** |
| **7** | 0.01 | 0.10 | **0.89** | 0.00 | 0.14 | **0.86** | 0.00 | 0.08 | **0.92** |
| **8** | 0.01 | 0.10 | **0.89** | 0.00 | 0.14 | **0.86** | 0.00 | 0.08 | **0.92** |
| **9** | 0.01 | 0.10 | **0.89** | 0.00 | 0.14 | **0.86** | 0.00 | 0.09 | **0.91** |
| **10** | 0.01 | 0.10 | **0.89** | 0.00 | 0.14 | **0.86** | 0.02 | 0.08 | **0.90** |
| 11 | 0.05 | 0.20 | 0.75 | 0.00 | 0.16 | 0.84 | 0.00 | 0.17 | 0.83 |
| 12 | 0.05 | 0.20 | 0.75 | 0.05 | 0.13 | 0.82 | 0.00 | 0.23 | 0.77 |
| 13 | 0.32 | 0.12 | 0.56 | 0.07 | 0.13 | 0.80 | 0.00 | 0.41 | 0.58 |
| 14 | 0.05 | 0.41 | 0.54 | 0.00 | 0.23 | 0.77 | 0.25 | 0.53 | 0.22 |
| 15 | 0.24 | 0.50 | 0.26 | 0.13 | 0.13 | 0.74 | 0.47 | 0.38 | 0.16 |
| 16 | 0.74 | 0.01 | 0.25 | 0.00 | 0.28 | 0.72 | 0.00 | 0.84 | 0.15 |
| 17 | 0.88 | 0.03 | 0.09 | 0.00 | 0.30 | 0.70 | 0.01 | 0.83 | 0.15 |
| 18 | 0.88 | 0.03 | 0.09 | 0.00 | 0.30 | 0.70 | 0.00 | 0.87 | 0.13 |
| 19 | 0.88 | 0.03 | 0.09 | 0.15 | 0.17 | 0.69 | 0.25 | 0.63 | 0.12 |
| 20 | 0.88 | 0.03 | 0.09 | 0.02 | 0.32 | 0.66 | 0.00 | 0.89 | 0.11 |

*4.3 System design synthesis and evaluation*

The 10 final designs of each component are seen as ten discrete values of each component to sample their combinations. To evaluate the system level performance with final component designs, 20 combinations of the three components are generated and each combination is achieved by randomly sampling one from the ten final designs of each of three components (These 20 new system designs are numbered as **No. 1' to 20'**). One should not confuse these 20 new system designs with the original component design instances in Table 3 (numbered as **No. 1 to 20**). The 20 new "good" component design combinations are integrated back to the simplified vehicle model shown in Figure 5(b) to generate 20 system designs with updated parts $P_2$, $P_3$ and $P_4$. They are simulated to calculate the responses (i.e. *SEA* and *M*) at the system level. By comparing their response with the original design (model), the performance of the new design method can be demonstrated.

## 5. Results

*5.1 Component performance improvement*

The simulated responses (*SEA* and *M*) of the 20 design alternatives for each critical component listed in Table 3 are shown in Figure 17(a-c), where a minus sign is added to the magnitude of *SEA*, to ensure that the data points with better performance concentrate at the lower-left corner. For comparison purpose, the original design for each component is also included. The results show that the DTUT-generated designs exhibit better performance on the *SEA* response but make moderate improvement on the *M*. As one can see in Figure 17(a) for P2, SEA is increased by about 70% for all design alternatives while only a half have mass reduction. The result for $P_3$ (Figure 17(b)) shows a similar trend but a more scattered *SEA* distribution. It reduces the SEA by 59% in maximum. According to Figure 17(c), the reduction of *SEA* is up to 20%, which indicates a less degree of performance improvement in the new design.

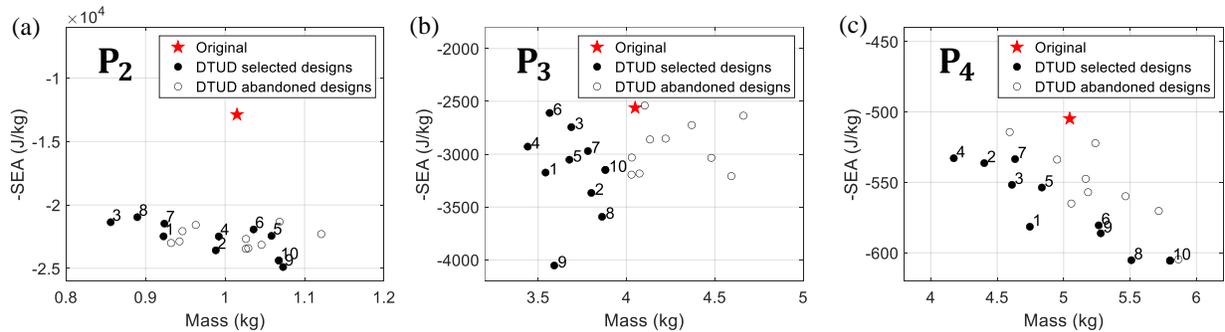

Figure 17 Comparing the basic vehicle model (the original design) with the twenty design alternatives generated by DTUD. 10 design alternatives screened by the DTUD under the uncertainty in design variables are denoted by the solid circles with the No. in Table 3 for (a) $P_2$, (b) $P_3$, and (c) $P_4$

After the screening process described in Section 4.2, the 10 cases with high reliability (i.e. No. 1~10 in Table 3) are highlighted. Much better performance of $M$ for $P_2$ and $P_3$, and both $M$ and $SEA$ for $P_4$ can be achieved.

*5.2 Final designs evaluation at the system level*

The 20 designs at the system level described in Section 4.3 (No. 1'~20') are simulated and the results are used to evaluate the effect of new component designs on the system level performance. Based on the simulation results, the distribution of total mass and vehicle structure intrusion are plotted in Figure 18, together with the data based on the original design (mass = 25.3 kg and intrusion = 238.1 mm). The comparison shows that most of the designs produced by the new DMM method perform better in terms of both responses. The maximum mass saving is 13.4% (21.9 kg), and the maximum intrusion reduction is 25.9% (176.5 mm). This indicates the performance improvement from the new design methodology.

In Figure 18, the intrusion and mass distributions of the 20 designs are approximated by normal distributions. It can be determined that the probabilities of the new design with lower intrusion and mass than the original design, are 94.3% and 85.7%, respectively. This again indicates the high performance of the new method.

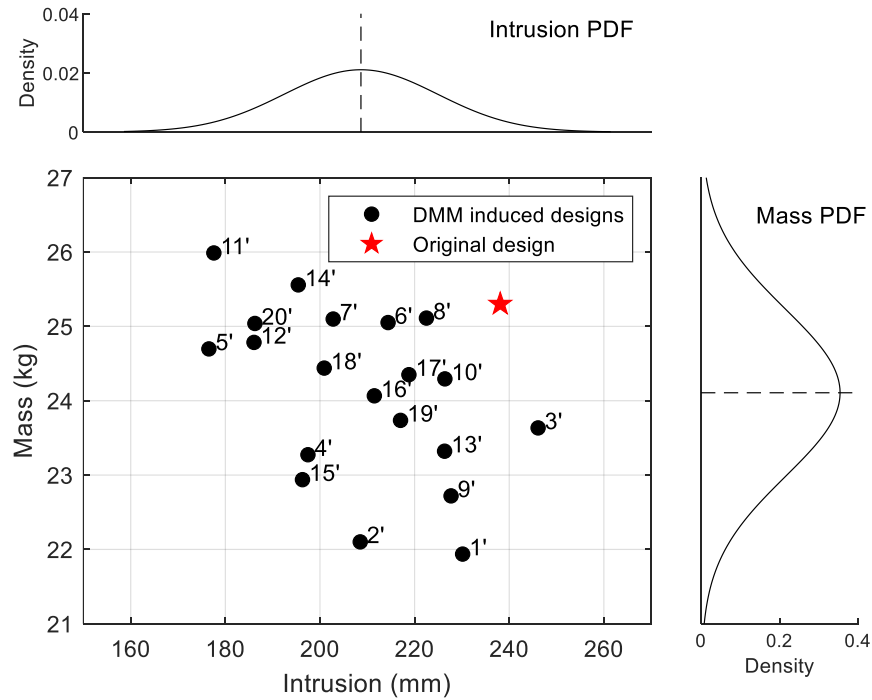

Figure 18 Distribution of the mass and intrusion for all 20 new designs and the results of the original design.

An additional comparison is made on the acceleration–time history at the location of the mass center of the vehicle model, as shown in Figure 19. The mean and standard deviation bounds of the 20 designs by DMM are also presented to analyze their response distribution. A lower acceleration (absolute value) indicates better performance. The results show that the acceleration of the original design and the DMM-generated designs exhibit a similar pattern. The peak acceleration occurs between 0.04 s and 0.06 s. Most of the 20 designs exhibit a lower peak acceleration than the original model. The magnitude of peak acceleration in the mean curve is 5% lower than that for the original design. The lower $\sigma$ bound is close to the peak of the original model so the new designs have a high probability to have a lower peak acceleration.

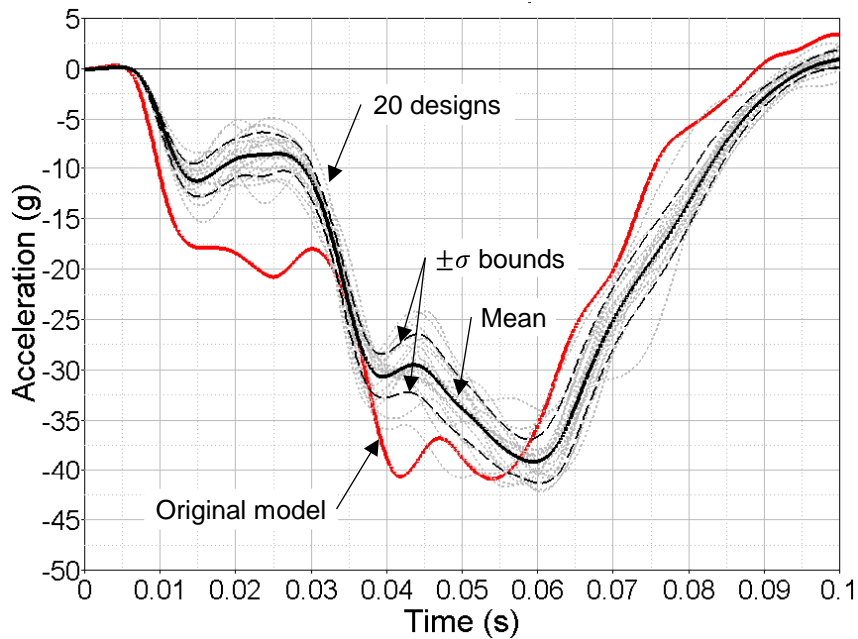

Figure 19 Comparison of acceleration histories between the original model and the 20 designs by DMM with the mean and standard deviation bounds ($\pm\sigma$ bounds) of the 20 designs

In an additional comparison, the firewall deformation modes are examined by comparing the original design with four randomly selected designs in Figure 20. The firewall is a nearly-2D structure separating the engine and passenger compartment, and its intrusion or degree of deformation is another indicator of vehicle crashworthiness. The deformation contours of the four selected designs when t = 90 ms are analyzed, together with the result for the original design. The comparison shows that the new designs can decrease the maximum deformation of the firewall by 20.3~35.6%. This indicates that the chance of the passenger's lower limb injury can be reduced significantly.

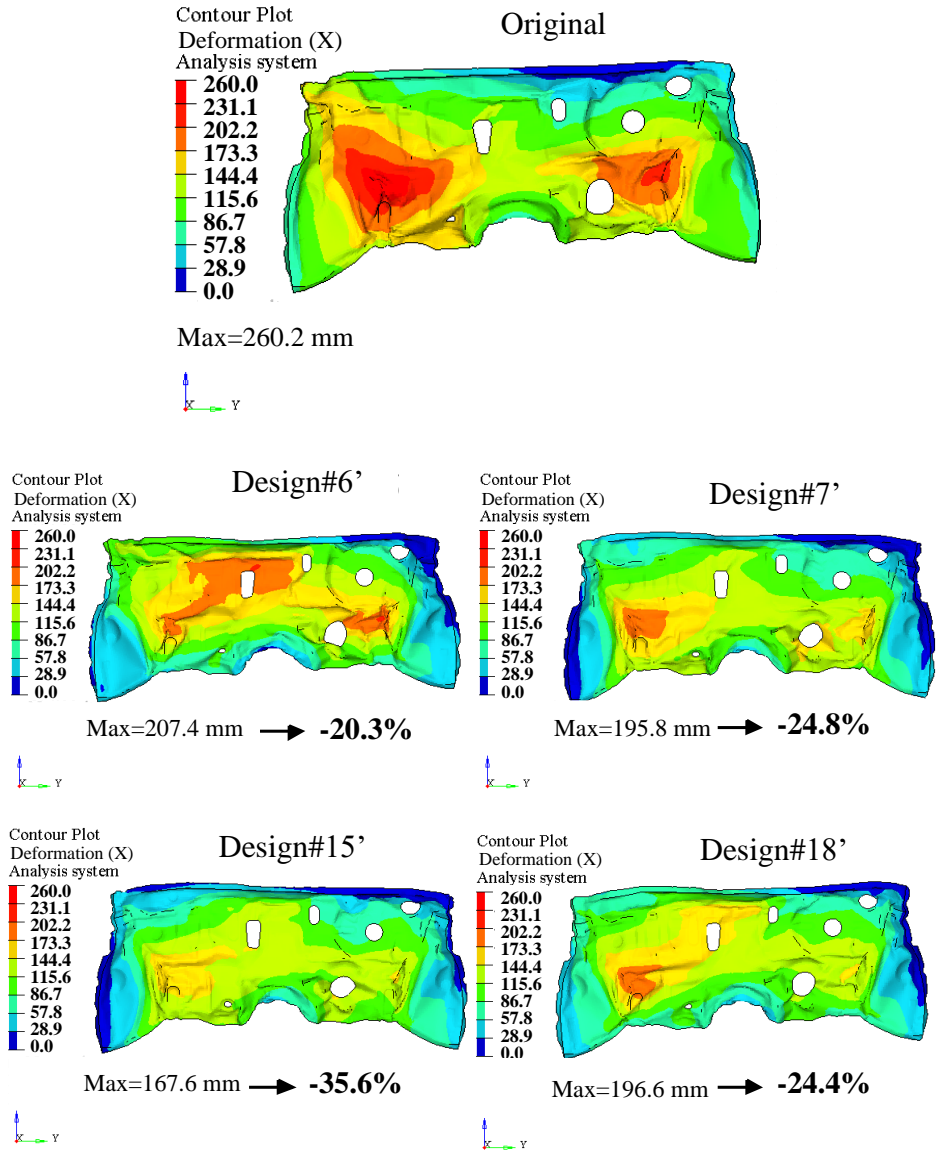

Figure 20 Comparison of the firewall (front view) deformation contours of the original design and four new designs at t = 90 ms. The maximum deformation magnitude is also shown together with the percentage decrease of maximum deformation compared to the original design.

## 6. Discussion

Discussions are made on the selection of representative 'g' branch based on the branch LP distribution and the branch-determined subspace volume. A decision tree may contain many 'g' branches and they need to be evaluated based on their accuracy and robustness to determine which one to be adopted. Besides the $ACC$, a new index, i.e., $CTT$ (Correctly classified 'g' tuples to the Total number of tuples) is used to quantify the robustness of the 'g' design subspace determined by a branch.

$CTT$ represents the relative size of the identified 'g' subspace to the initial design space. For a branch with a label $L_t$, $CTT$ can be calculated by

$$CTT = \frac{\sum_{i=1}^{n_{Lt}} p_i^{t_{Lt}}}{|D_o|}, \qquad (21)$$

where $|D_o|$ is the size of the training dataset. A larger $CTT$ value indicates the high capability of generated designs to reduce the impact of uncertainty in design variables.

In this study, the $ACC$ and $CTT$ values of all the 'g' branches on the three DTUDs at the component design are plotted in Figure 21, where the minus values are used to locate the preferred values at the lower-left corner. In Figure 21, for $P_2$ and $P_3$, the branches with the high $CTT$ and good $ACC$ (about 90%) are selected, i.e., b3 and b19, respectively. The b19 is selected for $P_4$ due to its relatively higher $ACC$ compared with b8 and b29, despite its slightly lower $CTT$.

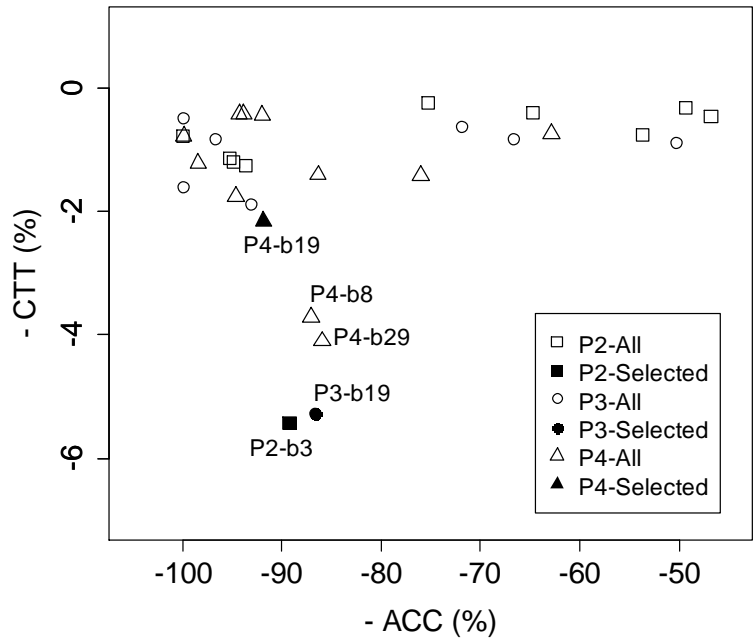

Figure 21 $ACC$ and $CTT$ for the 'g' branches of the DTUDs of the three critical components ($P_2$, $P_3$, and $P_4$)

In addition, the decision tree complexity should be considered for applying DMM in design. The high complexity often increases modeling accuracy but may cause overfitting and low interpretability. For a specific dataset or design problem, a model with optimal complexity exists with the minimum generalization error [34]. Many strategies have been used for complexity control [35]. In this study, the number of decision tree layers is controlled for good interpretability. Generally, to better interpret design problems, fewer layers are preferred, for example, the six-layer decision tree at the system level is shown in Figure 10. More layers can be helpful to reach a subspace with better $ACC$, for example, the nine-layer DTUD at the component level.

## 7. Conclusions

In this study, a data mining method (DMM) based on the decision tree technique is developed for vehicle crashworthiness design. A new decision tree algorithm for uncertain data (DTUD) has been developed by extending the traditional decision tree (TDT) algorithm and considering the uncertainty in design variables. The framework of this new design method is implemented with three main steps, that is, the design at the system level, design at the component level, and evaluation of the whole vehicle performance. At the system level, the dataset is generated by combining the DOE and FE simulations. Decision tree is constructed to derive design rules, i.e., the critical component, their performance requirements, and boundary conditions. At the component level, the identified critical components are designed based on a new DTUD by incorporating the rules from the system level. The performance variations under uncertainty in design variables can be controlled by DTUD. The newly designed components are integrated with the vehicle model to verify the performance.

As a case study, the new method is applied to a simplified and validated 2010 Toyota Yaris passenger car model to improve its crashworthiness. A large number of simulations are conducted on the DOEs and the results form a design dataset. A TDT is trained from the dataset and used to identify the key energy-absorbing components and determine the design subspaces and boundary conditions for each key component.

The information obtained at the system level is then used for detailed component level design. By learning from the dataset with uncertainty in design variables, the DTUDs at the component level are built for each key component. The critical geometric design variables and their ranges are then revealed by the DTUD. Twenty design alternatives are generated based on these rules for each critical component, and their response is verified by FE simulations. After screening out ten low-reliability samples by the trained DTUD, the response performance and its reliability under uncertainty are further improved.

The newly design components are then integrated back to the simplified vehicle model and crash simulations are conducted. The results demonstrate that the new designs outperform the original design in terms of mass, intrusion, and peak acceleration. The maximum decreases of these responses are 13.4%, 12.5%, and 35.6%, respectively. The performance of the new design method is verified and can potentially be applied to other complex and hierarchical systems.

**Acknowledgments**

The authors would like to thank Dr. Hongyi Xu at University of Connecticut for his valuable suggestions during the course of this work and Dr. Jun Wu at Hunan Normal University for his help in morphing modeling.

**Appendix A**

*Morphing method for the geometry shape control*

In Appendix A, a simple surface modeled with thin shell elements shown in Figure A.1 is used as an example to describe the mesh morphing method. The 3D surface is parameterized by five Morphing Control points (MCPs), where each MCP has three coordinates ($x$, $y$, and $z$). By changing MCP#5, the model geometry is changed accordingly by this morphing method.

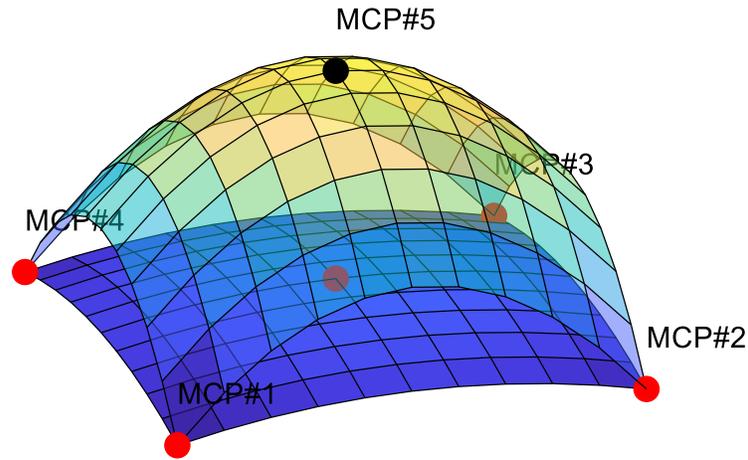

Figure A.1 Shell element surface before and after the morphing operation with one of the Morphing Control Point (MCP#5) changed

In the morphing algorithm, the radial basis function is used to fit the rule of morphing original control points to their new locations with Eq. (A.1)

$$[x \quad y \quad z] = \Phi(r_{ij})\chi + [X \quad Y \quad Z]k + c \tag{A.1}$$

where $[x \quad y \quad z]$ and $[X \quad Y \quad Z]$ are the matrices of the new and original control point matrices, respectively. $\Phi(r_{ij}) = r_{ij}^2 \log(r_{ij})$ is the basis function, where $r_{ij}$ is the Euclidean distance between the $i$th original to the $j$th new MCPs. In the matrix form, it can be written as

$$\begin{bmatrix} A & B \\ B^T & 0 \end{bmatrix} \begin{bmatrix} \chi \\ [k] \\ c \end{bmatrix} = \begin{bmatrix} M_{cp} \\ 0 \end{bmatrix} \tag{A.2}$$

where $A$ is the basis function matrix with $A_{ij} = \Phi(r_{ij})$. $B = \begin{bmatrix} 1 & X_1 & Y_1 & Z_1 \\ \vdots & \vdots & \vdots & \vdots \\ 1 & X_n & Y_n & Z_n \end{bmatrix}_{n \times 4}$ with X, Y, and Z being the coordinates of the original MCPs. $M_{cp} = \begin{bmatrix} x_1 & y_1 & z_1 \\ \vdots & \vdots & \vdots \\ x_n & y_n & z_n \end{bmatrix}_{n \times 3}$ is the coordinate matrix of the new

MCPs. The morphing rule ($\begin{bmatrix} \chi \\ [k] \\ c \end{bmatrix}$) can be determined by solving Eq. (A.2). This rule is used for the morphing of the FE nodes from the baseline FE model to the new one based on the new location of MCPs by Eq. (A.3)

$$\begin{bmatrix} A' & B' \\ B'^T & 0 \end{bmatrix} \begin{bmatrix} \chi \\ [k] \\ c \end{bmatrix} = \begin{bmatrix} M_{nd} \\ 0 \end{bmatrix} \quad (A.3)$$

where $A'_{ij} = \Phi(r'_{ij})$ with $r'_{ij}$ is the Euclidean distance between the $i$th basic FE node and $j$th original MCP. $B = \begin{bmatrix} 1 & X_1 & Y_1 & Z_1 \\ \vdots & \vdots & \vdots & \vdots \\ 1 & X_m & Y_m & Z_m \end{bmatrix}_{m \times 4}$, where $m$ is the number of basic FE model nodes. $M_{nd} = \begin{bmatrix} x_1 & y_1 & z_1 \\ \vdots & \vdots & \vdots \\ x_m & y_m & z_m \end{bmatrix}_{m \times 3}$ is the coordinate matrix of the new FE model nodes. In this way, the projection rule between the original MCPs and the new MCPs is fitted by the $\begin{bmatrix} \chi \\ [k] \\ c \end{bmatrix}$, which is then used to morph the basic FE model (nodes) to a new model, that is, the new FE nodes' locations. This algorithm is implemented as a MATLAB script to realize an automatic transformation.

# Appendix B

*Uncertainty evaluation for the labels of design alternatives*

*Table B.1 The 20 design alternatives generated by the b3 of DTUD for $P_2$ and the predicted LPs. The 10 designs with high 'g' LP (bold) are selected as the final designs.*

|     | Design variables ||||||||| | LP ("g" :0.89) |||
| --- | --- | --- | --- | --- | --- | --- | --- | --- | --- | --- | --- | --- | --- |
| NO. | YN1 | YN2 | YN3 | YN4 | ZN1 | ZN2 | ZN3 | ZN4 | T2 | T3 | p | m | g |
| **1** | **2.44** | **-13.79** | **10.31** | **-6.53** | **3.74** | **-18.30** | **11.77** | **-19.51** | **2.08** | **1.40** | **0.05** | **0.00** | **0.95** |
| **2** | **-13.26** | **-9.48** | **7.93** | **2.43** | **-4.43** | **-6.55** | **12.03** | **0.38** | **2.04** | **1.24** | **0.05** | **0.00** | **0.95** |
| **3** | **-14.43** | **-6.67** | **6.05** | **-8.69** | **-1.61** | **-7.90** | **7.32** | **1.63** | **1.52** | **1.63** | **0.04** | **0.01** | **0.95** |
| **4** | **-7.31** | **-10.35** | **-7.82** | **-0.88** | **-0.70** | **-12.31** | **19.32** | **-9.21** | **2.27** | **1.02** | **0.00** | **0.06** | **0.94** |
| **5** | **-4.35** | **-7.32** | **5.34** | **-14.33** | **19.43** | **1.21** | **15.93** | **3.24** | **1.87** | **1.23** | **0.01** | **0.10** | **0.89** |
| **6** | **-1.85** | **-9.57** | **-8.51** | **-11.60** | **10.96** | **-19.32** | **-8.88** | **-15.53** | **1.79** | **1.58** | **0.01** | **0.10** | **0.89** |
| **7** | **0.08** | **-8.12** | **-1.62** | **-14.75** | **-3.81** | **0.29** | **-0.14** | **1.02** | **2.11** | **1.12** | **0.01** | **0.10** | **0.89** |
| **8** | **-7.07** | **-14.78** | **0.37** | **-8.90** | **9.30** | **-17.18** | **3.51** | **-7.00** | **1.69** | **1.30** | **0.01** | **0.10** | **0.89** |
| **9** | **-6.14** | **-12.94** | **-10.25** | **-7.27** | **17.56** | **1.99** | **-17.03** | **7.18** | **1.99** | **1.07** | **0.01** | **0.10** | **0.89** |
| **10** | **-10.58** | **-9.89** | **-0.34** | **-13.66** | **18.33** | **-11.78** | **17.05** | **16.14** | **1.93** | **1.52** | **0.01** | **0.10** | **0.89** |
| 11 | -0.04 | -11.98 | 2.16 | -6.30 | -16.42 | -14.46 | 13.47 | -16.02 | 1.72 | 1.59 | 0.05 | 0.20 | 0.75 |
| 12 | -8.47 | -11.46 | 1.64 | -1.89 | -15.08 | 2.44 | -7.62 | 8.08 | 2.00 | 1.08 | 0.05 | 0.20 | 0.75 |
| 13 | -12.64 | -8.29 | 5.74 | -1.38 | -10.31 | -5.08 | -4.00 | -1.23 | 2.18 | 1.18 | 0.32 | 0.12 | 0.56 |
| 14 | -6.34 | -13.31 | 14.91 | -5.77 | -4.82 | -18.68 | -13.97 | 10.64 | 1.96 | 1.61 | 0.05 | 0.41 | 0.54 |
| 15 | -3.30 | -12.32 | -7.73 | -13.18 | -15.38 | -15.58 | -13.37 | 7.63 | 1.83 | 1.65 | 0.24 | 0.50 | 0.26 |
| 16 | -3.78 | -11.90 | -3.75 | -6.95 | -13.39 | 6.60 | 4.97 | -0.31 | 2.26 | 1.28 | 0.74 | 0.01 | 0.25 |
| 17 | -5.63 | -13.50 | -6.12 | -9.80 | 13.50 | 14.05 | 8.07 | 11.35 | 1.66 | 1.97 | 0.88 | 0.03 | 0.09 |
| 18 | -9.16 | -14.29 | -3.46 | -8.25 | -5.97 | 19.31 | -2.90 | 18.29 | 2.29 | 1.12 | 0.88 | 0.03 | 0.09 |
| 19 | 5.38 | -9.13 | -4.85 | 1.98 | 13.96 | 12.42 | 2.47 | -17.48 | 1.71 | 1.36 | 0.88 | 0.03 | 0.09 |
| 20 | 3.81 | -9.93 | -9.72 | 1.19 | -10.57 | 17.88 | 1.85 | 13.63 | 2.05 | 1.52 | 0.88 | 0.03 | 0.09 |

*Table B.2 The 20 design alternatives generated by the b19 of DTUD for Part $P_3$ and the predicted LPs. The 10 designs with high 'g' LP (bold) are selected as the final designs.*

|     | Design variables ||||||||| | LP ("g": 0.87) |||
| --- | --- | --- | --- | --- | --- | --- | --- | --- | --- | --- | --- | --- | --- |
| NO. | YN1 | YN2 | YN3 | YN4 | ZN1 | ZN3 | ZN4 | d | T4 | T5 | p | m | g |
| **1** | **-14.81** | **-10.31** | **-3.31** | **-9.35** | **-0.13** | **-17.03** | **1.36** | **0.32** | **1.69** | **1.51** | **0.00** | **0.13** | **0.87** |
| **2** | **-2.03** | **-11.48** | **-11.64** | **9.70** | **2.71** | **-5.26** | **3.34** | **1.91** | **1.52** | **1.92** | **0.00** | **0.13** | **0.87** |
| **3** | **2.63** | **5.98** | **-2.92** | **4.65** | **5.31** | **-16.06** | **-6.74** | **4.94** | **1.62** | **1.63** | **0.00** | **0.13** | **0.87** |
| **4** | **-9.47** | **-6.84** | **2.67** | **9.24** | **-17.11** | **2.65** | **-7.98** | **2.71** | **1.55** | **1.55** | **0.00** | **0.13** | **0.87** |
| **5** | **-11.03** | **8.71** | **-14.36** | **2.02** | **-1.92** | **-8.12** | **18.89** | **2.60** | **1.62** | **1.61** | **0.00** | **0.13** | **0.87** |
| **6** | **2.14** | **-5.12** | **2.38** | **-10.26** | **15.11** | **7.40** | **-8.44** | **2.47** | **1.58** | **1.53** | **0.00** | **0.13** | **0.87** |
| **7** | **-11.83** | **3.96** | **7.71** | **12.96** | **3.44** | **0.30** | **1.68** | **4.75** | **1.72** | **1.56** | **0.00** | **0.14** | **0.86** |
| 8 | 1.31 | -3.00 | -11.18 | -10.81 | -1.52 | -2.15 | -1.14 | 4.03 | 1.71 | 1.77 | 0.00 | 0.14 | 0.86 |
| 9 | -14.01 | 11.46 | -13.38 | -3.26 | 9.74 | 0.87 | -4.00 | 4.43 | 1.53 | 1.65 | 0.00 | 0.14 | 0.86 |
| 10 | 0.72 | 5.22 | -6.63 | -11.58 | -2.76 | 2.18 | 6.04 | 3.21 | 1.74 | 1.70 | 0.00 | 0.14 | 0.86 |
| 11 | -7.81 | -5.75 | 4.58 | -13.67 | -4.21 | -7.22 | 17.37 | 2.27 | 1.70 | 2.45 | 0.00 | 0.16 | 0.84 |
| 12 | -10.13 | -0.58 | -1.87 | -6.65 | -19.99 | 16.81 | -15.03 | 1.22 | 1.72 | 2.15 | 0.05 | 0.13 | 0.82 |
| 13 | 4.98 | 1.72 | -0.98 | -0.85 | -3.37 | 8.61 | -13.97 | 1.64 | 1.73 | 2.28 | 0.07 | 0.13 | 0.80 |
| 14 | -10.60 | 10.24 | 0.27 | 8.81 | 17.57 | -3.34 | -19.86 | -0.01 | 1.77 | 1.86 | 0.00 | 0.23 | 0.77 |
| 15 | 4.54 | 0.97 | -8.51 | 13.24 | -12.61 | 3.35 | -15.34 | 3.76 | 1.74 | 2.12 | 0.13 | 0.13 | 0.74 |
| 16 | -0.99 | 5.79 | -5.58 | -14.98 | 18.22 | 4.47 | 15.39 | 1.77 | 1.79 | 1.71 | 0.00 | 0.28 | 0.72 |
| 17 | -11.34 | -8.57 | -1.58 | 7.50 | -17.94 | 10.31 | 11.72 | 0.96 | 1.80 | 1.83 | 0.00 | 0.30 | 0.70 |
| 18 | -14.48 | -5.93 | 4.29 | -14.21 | 1.88 | -12.72 | -17.03 | 4.63 | 1.78 | 1.97 | 0.00 | 0.30 | 0.70 |
| 19 | -13.26 | 3.10 | 8.52 | 7.85 | 6.57 | -15.36 | 14.78 | 2.97 | 1.75 | 2.42 | 0.15 | 0.17 | 0.69 |
| 20 | 1.50 | 11.90 | 3.59 | -2.00 | 14.26 | 8.95 | -3.23 | 1.97 | 1.70 | 2.49 | 0.02 | 0.32 | 0.66 |

*Table B.3 The 20 design alternatives generated by the b19 of DTUD for Part $P_4$ and the predicted LPs. The 10 designs with high 'g' LP (bold) are selected as the final designs.*

| | Design variables | | | | | | | | | | | LP ("g": 0.92) | | |
|---|---|---|---|---|---|---|---|---|---|---|---|---|---|---|
| NO. | YN1 | YN2 | YN3 | YN4 | ZN1 | ZN2 | ZN3 | ZN4 | T6 | T7 | T8 | p | m | g |
| **1** | **2.82** | **-11.41** | **-10.32** | **-7.94** | **8.90** | **-10.95** | **14.63** | **13.31** | **2.75** | **2.49** | **1.53** | **0.00** | **0.06** | **0.94** |
| **2** | **-5.22** | **9.87** | **-4.70** | **-7.61** | **8.96** | **-13.86** | **9.34** | **-4.49** | **2.05** | **2.65** | **1.56** | **0.00** | **0.08** | **0.92** |
| **3** | **-3.21** | **9.19** | **-2.87** | **0.30** | **1.26** | **-12.17** | **12.52** | **6.83** | **2.43** | **2.19** | **1.51** | **0.00** | **0.08** | **0.92** |
| **4** | **-12.98** | **-14.37** | **10.91** | **-12.71** | **14.67** | **-14.59** | **13.13** | **-13.16** | **2.21** | **2.70** | **1.51** | **0.00** | **0.08** | **0.92** |
| **5** | **10.25** | **-12.49** | **-6.38** | **-8.82** | **0.83** | **-13.55** | **10.55** | **7.57** | **2.27** | **2.35** | **1.91** | **0.00** | **0.08** | **0.92** |
| **6** | **4.39** | **14.48** | **-8.73** | **-14.10** | **9.97** | **-14.11** | **6.16** | **4.57** | **2.81** | **2.30** | **1.73** | **0.00** | **0.08** | **0.92** |
| **7** | **-13.40** | **-0.02** | **-2.26** | **4.25** | **6.60** | **-12.43** | **1.50** | **7.40** | **2.17** | **2.70** | **1.77** | **0.00** | **0.08** | **0.92** |
| **8** | **-14.73** | **6.53** | **-13.09** | **12.37** | **8.10** | **-13.51** | **13.99** | **-2.30** | **2.98** | **2.33** | **1.94** | **0.00** | **0.08** | **0.92** |
| **9** | **3.65** | **-1.48** | **-5.37** | **-11.28** | **3.95** | **-14.36** | **2.78** | **8.16** | **2.64** | **2.80** | **1.96** | **0.00** | **0.09** | **0.91** |
| **10** | **1.36** | **7.55** | **5.91** | **5.24** | **1.38** | **-14.95** | **10.20** | **-10.38** | **2.61** | **2.99** | **1.99** | **0.02** | **0.08** | **0.90** |
| 11 | -11.56 | -7.29 | -1.04 | -6.66 | 12.38 | -12.88 | 5.86 | -11.17 | 2.67 | 2.09 | 2.03 | 0.00 | 0.17 | 0.83 |
| 12 | 6.26 | -6.06 | 12.33 | -10.39 | 2.19 | -12.95 | 3.62 | 9.51 | 2.92 | 2.97 | 2.05 | 0.00 | 0.23 | 0.77 |
| 13 | -10.59 | -4.90 | -0.24 | 1.87 | 11.38 | -14.48 | 6.65 | 11.94 | 2.08 | 2.44 | 2.10 | 0.00 | 0.41 | 0.58 |
| 14 | 11.05 | 10.51 | -9.17 | -6.36 | 2.96 | -11.64 | 8.90 | -12.45 | 2.95 | 2.63 | 1.57 | 0.25 | 0.53 | 0.22 |
| 15 | -7.14 | 14.32 | 6.43 | -13.76 | 11.85 | -11.00 | 7.05 | -3.33 | 2.10 | 2.11 | 2.08 | 0.47 | 0.38 | 0.16 |
| 16 | 2.22 | 8.48 | 13.07 | -1.83 | 5.81 | -11.44 | 0.47 | -10.62 | 2.46 | 2.77 | 1.81 | 0.00 | 0.84 | 0.15 |
| 17 | 9.82 | 1.96 | 4.38 | 5.56 | 8.36 | -10.71 | 9.74 | 5.25 | 2.37 | 2.87 | 1.97 | 0.01 | 0.83 | 0.15 |
| 18 | 0.01 | 13.51 | -9.73 | 1.44 | 9.29 | -11.72 | 12.26 | 3.47 | 2.78 | 2.12 | 2.06 | 0.00 | 0.87 | 0.13 |
| 19 | -8.28 | 1.17 | 2.89 | 6.99 | 14.98 | -11.74 | 4.15 | 6.31 | 2.36 | 2.54 | 2.07 | 0.25 | 0.63 | 0.12 |
| 20 | -4.27 | -13.34 | -5.66 | 4.06 | 1.75 | -11.37 | 4.68 | -0.20 | 2.33 | 2.74 | 1.80 | 0.00 | 0.89 | 0.11 |